\documentclass[12pt,preprint]{aastex}

\slugcomment{ }

\shorttitle{Chemistry and PAH's in dense clouds}
\shortauthors{Wakelam \& Herbst}

\begin{document}

\title{Polycyclic Aromatic Hydrocarbons (PAH's) in dense cloud chemistry}

\author{V. Wakelam}
\affil{Universit\'e Bordeaux 1, CNRS, LAB , BP89 33270 Floirac, France }
\and
\author{E. Herbst}
\affil{Departments of Physics,  Astronomy and Chemistry, The Ohio State University, Columbus, OH 43210, USA}




\begin{abstract}
Virtually all detailed gas-phase models of the chemistry of dense interstellar clouds exclude  polycyclic aromatic hydrocarbons (PAH's).  This omission is unfortunate because from the few studies that have been done on the subject, it is known that the inclusion of PAH's can affect the gas-phase chemistry strongly. We have added  PAH's to our network to determine the role they play in the chemistry of cold dense cores. Initially, only the chemistry of neutral and negatively-charged PAH species was considered, since it was assumed that positively-charged PAH's  are of little importance. Subsequently, this assumption was checked and confirmed.  In the models presented here, we include radiative attachment to form PAH$^{-}$, mutual neutralization between PAH anions and small positively-charged ions, and photodetachment. We also test the sensitivity of our results to changes in the size and abundance of the PAH's. Our results confirm that the inclusion of PAH's changes many of the calculated abundances of smaller species considerably. In TMC-1, the general agreement with observations is significantly improved contrary to L134N. This may indicate a difference in PAH properties between the two regions. With the inclusion of PAH's in dense cloud chemistry,  high-metal elemental abundances  give a satisfactory agreement with observations. As a result, we do not need to decrease the observed elemental abundances of all metals  and we do not need to vary the elemental C/O ratio in order to produce large abundances of carbon species in TMC-1 (CP). 
\end{abstract}




%

\section{Introduction}

Polycyclic Aromatic Hydrocarbons (PAH's) are hydrocarbon compounds with multiple six-membered aromatic rings.  Such species are thought to be a common component of the interstellar medium in the Milky Way and in external galaxies \citep{2003ARA&A..41..241D}. PAH's are typically observed in infra-red emission bands following excitation by either ultra-violet or visible radiation in assorted regions of the interstellar medium exposed to radiation \citep{1985A&A...146...81L,1989A&A...216..148L,1985ApJ...290L..25A,1999ApJ...526..265S}. Despite all attempts, no specific types of PAH's responsible for these features have been identified \citep{2006A&A...456..161M}, and the form and intensity of the spectra vary strongly from one region to another \citep{2006astro.ph..8003D}, probably reflecting a variety of sizes and abundances. The origin of PAH's also remains an unanswered question \citep{2003A&G....44f..14W}. Most believe that PAH's are formed in the atmospheres of carbon stars \citep{1997A&A...323..163A,2002A&A...392..203C} or in supernovae \citep{2006ApJ...653..267T}. 

Various observations show that the abundance of small free-flying PAH's have a tendency to decrease in dense clouds where UV photons do not penetrate \citep{1990ApJ...364..136B,1993A&A...277..609B,2005A&A...429..193R}. \citet{2005A&A...429..193R,2006astro.ph..9186R} proposed that PAH's form larger PAH's or PAH-clusters in such dense regions. As a consequence, free-flying PAH's observed at the border of illuminated clouds may be the result of photodissociation of these larger species. Although they determined a minimal size of 400 carbons for these species using mid-infrared ISOCAM maps of photodissociation regions (PDR's), this number is very uncertain.  If free-flying PAH's are not present in the dense interstellar medium,  an efficient process is needed to aggregate them.  Aggregation is poorly understood,  and the time in which this process occurs and  the  fraction of small PAH's remaining in dense clouds unexposed to intense radiation are undetermined. Currently, constraints on the PAH size and abundance as a function of the physical conditions in the dense interstellar medium remain weak.  In this article, we focus not on the size of PAH's but on the chemical impact of their presence in the dense cold interstellar cores.  Nevertheless, the sensitivity of our results to size will be examined.   We follow past authors by lumping all PAH's  into  three groups of species: neutral, positively charged, and negatively charged PAH's, designated PAH, PAH$^{+}$,  and PAH$^{-}$ respectively. Of these three groupings, PAH$^{+}$ species were found to be unimportant and were excluded from the detailed calculations.

The chemical effect of PAH's at steady-state has been studied previously.  In the diffuse interstellar medium, the inclusion of PAH's, among other effects,  enhances the neutralization of  atomic ions such as carbon, silicon and sulfur, reducing their abundances \citep{1988ApJ...329..418L,1998ApJ...499..258B}.  In their study of dense cloud chemistry, \citet{1988ApJ...324..553L} concluded that for PAH's consisting of molecules with 30-50 carbon atoms, with a fractional abundances greater than $10^{-8}$, PAH$^-$ is the main source of negative charge, which considerably changes the abundances of some neutral species, enhancing, for example, the abundances of carbon-bearing molecules. Finally, \citet{2003MNRAS.343..390F} studied the importance of PAH's for C-shock models.  In this article, we report a time-dependent study of the role of PAH's in the gas-phase chemistry of cold dense clouds and compare our results with observations.  One focus of our study will be how the presence of PAH's affects the choice of elemental abundances.

The elemental abundances typically used for chemical models of dense clouds are referred to by the appellation ``low-metal''.  Originally defined by \citet{1982ApJS...48..321G}, these abundances are based on the  {\it Copernicus} satellite observations of the $\zeta$ Ophiuchi diffuse region \citep{1974ApJ...193L..35M}. But, in order to reproduce the observation of molecules in dense clouds better, the $\zeta$ Oph  abundances of all the elements but, He, C, N and O are decreased by two orders of magnitude.  This assumption implies that a strong depletion of the heavier elements occurs between the diffuse and dense cloud phases \citep{1982ApJS...48..321G}. Although the details of this depletion are not understood, the depleted abundances are necessary to reproduce  the abundances of molecular ions and, to a lesser extent, carbon-bearing species.   Since the presence of PAH's in dense clouds should modify the ionization fraction, it may help in solving the problem of the choice of elemental abundances.

This article is organized as follows. In Sect.~\ref{PAH_model}, we describe the parameters of the PAH's used for the analysis and the chemical processes involving them.  In Sect.~\ref{model_other}, other details of the chemical model are given. The influence of PAH's on the abundances of smaller species is reported in the next section, while the sensitivity of the results to the PAH parameters is discussed in Sect.~\ref{sensitivity}.  In Sect.~6, we compare observations in molecular clouds with grids of models in which we vary several parameters, including PAH size and abundance. The paper ends with our conclusions.

\section{PAH parameters and chemistry}\label{PAH_model}

\citet{2006astro.ph..8003D} reproduced the extinction curves of various diffuse regions of the Milky Way from {\it Spitzer} observations and constrained the size distribution of carbonaceous grains including PAH's down to 3$\AA$ in size \citep[see Fig.~6 of][]{2006astro.ph..8003D}. Using the model $j_M=1$ of \citet{2006astro.ph..8003D}, which has the largest abundance of small PAH's, we integrated over a radius range of 3.8 to 6.0$\AA$\footnote{The dependence of the number of carbon atoms $N_{C}$  on radius $a$ for PAH's is not well-known since it depends on the structure of the molecule \citep{1986A&A...164..159O}. We used the relation given by \citet{2006astro.ph..8003D}: $N_C = 460 (a/10 \AA)^3$. The radius range here then corresponds  to PAH's with between 25 and 100 carbon atoms.}. The integration yields a mean radius of 4~$\AA$ ($N_C = 30$) and an integrated fractional abundance with respect to $n_{\rm H}$ of $3.07\times 10^{-7}$.  Multiplying by 30 leads to a fractional abundance of C atoms in PAH's of $9.1 \times 10^{-6}$, or roughly 3\% of the elemental carbon abundance. 

Because PAH's may not be free in dense clouds,  we decided to also consider  PAH-clusters of radius 10 and 100 $\AA$. 
With the formula of \citet{2006astro.ph..8003D}, these limits correspond to PAH's with 468 and $4.68\times 10^5$ carbon atoms, leading to fractional abundances of  $2\times 10^{-8}$ and $2\times 10^{-11}$, respectively, if one assumes (a) the same atomic carbon abundance in PAH's as for the smaller species, and (b) that weakly-bound clusters can also be approximated by the same relationship between number of carbon atoms and size. 
 
  \citet{1998ApJ...499..258B} and \citet{1997A&A...323..534D} found that the abundance of PAH$^+$ decreases strongly in clouds with an A$_v$ larger than 2 and \citet{1988ApJ...324..553L} found that the abundance of PAH$^+$ is smaller than the abundance of the other forms of PAH's by about one order of magnitude in their calculations. For this reason, in our study of dense cloud chemistry, we did not originally include cations but only neutral and negatively-charged PAH's. Prodded by the referee, however, we looked more carefully at the role of PAH$^+$, but concluded again that their exclusion is justified.  The reconsideration  will  be discussed in the next section.  We start the chemistry with neutral PAH's only,  based on the results of  \citet{1998ApJ...499..258B} and \citet{1997A&A...323..534D}. 
Neutral PAH's become negatively charged very rapidly and our results after a very small time interval do not depend on whether the chemistry commences with PAH or PAH$^-$.

\subsection{Chemical processes involving PAH's}\label{PAH_chem}

The main problem of including PAH's in chemical models is that few experimental measurements of reaction rates have been performed on these molecules, and these few tend to involve rather small cationic PAH's only.  For example, \citet{1998Natur.391..259S} studied the rates of the reactions between a variety of positively-charged PAH's with six, ten, and sixteen carbon atoms and the neutrals  H, O, N, and H$_{2}$, while 
 \citet{2006ApJ...651L.129B} studied reactions of coronene cation (C$_{24}$H$_{12}^+$) with the same reactants. Although association reactions appear to be dominant, most the experiments were done at sufficiently high densities that the mechanism, three-body association, is not what occurs in the low-density interstellar medium.  Moreover, cationic PAH's are not important in our study.
 
  The processes  in our model include electron attachment, recombination of negative PAH ions with smaller positive ions, and photodetachment of negative PAH ions by UV radiation.  We consider these in turn.

Negative PAH ions are formed by radiative electron attachment :
\begin{equation}
\rm PAH + e^-  \longrightarrow PAH^-  + h\nu.
\end{equation}
The rate coefficient for this reaction is thought to depend on the size of the PAH. \citet{1986A&A...164..159O} and \citet{1989ApJS...71..733A} proposed two different relations: $k_e = 10^{-7} (N_C)^{3/4}$~cm$^3$~s$^{-1}$ and $ k_e = 1.2\times 10^{-7} (N_C)^{1/2}$~cm$^3$~s$^{-1}$.  These two formulae lead to differences of about a factor of 2 for PAH's with $N_C < 100$. We arbitrarily decided to follow \citet{2001ApJS..132..233L} and use the first formula. The rate coefficient is $1.3\times 10^{-6}$~cm$^3$~s$^{-1}$ for the attachment of electrons to neutral PAH's with 30 carbons. The results depend on this value only if $ k_{\rm e}$ is decreased by a factor of 100. In other words, an increase of this rate coefficient does not change the results.   Calculations by Herbst (in preparation)  based on the statistical approach of \citet{Terzieva2000} indicate that with typical electron affinities in the range of 0.5 - 1.0 eV, the large rate coefficients deduced by previous authors are valid for neutral PAH's with 30 carbon atoms, but not for those significantly smaller, in agreement with the experimental results of \citet{1998cpmg.conf...71M}.

Negatively-charged PAH's can recombine with positive ions, designated X$^+$.  These recombination reactions differ from dissociative recombination of positive ions with electrons in two important ways \citep{1988rcia.conf...41B}.  First, they are less exothermic because of the electron affinity of the PAH.  Secondly, some of the reaction exothermicity can be deposited in internal energy of the PAH so that less is available to dissociate the positive ion upon neutralization.  We assume that two resulting channels are possible for these recombinations, simple charge exchange:
\begin{equation}\label{channel1}
\rm PAH^- + X^+   \longrightarrow  PAH + X,
\end{equation}
and, if the positive ion contains hydrogen: 
\begin{equation}\label{channel2}
\rm PAH^- + X^+   \longrightarrow  PAH + X' + H,
\end{equation}
where X' designates a neutral with one fewer H atom than the ion.  The terms ``neutralization'' and ``dissociative neutralization'' are used for these two processes to distinguish them from recombination with electrons.  We assume equal branching fractions for the two sets of products, if they exist. If the neutralization leads to a neutral species that does not exist or is not included in the model, only the second channel is considered. The products of the dissociation are very uncertain, and laboratory experiments are required. 

The rate coefficients for these reactions have been computed by taking into account the Coulomb attraction between the charged species via the expression \citep{1941ApJ....93..369S,1978MNRAS.184..227W}: 
\begin{equation}\label{k_grain_mol}
k  = \pi a^2 \left <v_{0}\right > \left (1+\frac{e^{2}}{a k_BT} \right ), 
\end{equation}
where $e$ is the electronic charge, $ k_B$ is the Boltzmann constant, $a$ is the radius of the PAH (or PAH cluster), and $ \left <v_{0}\right > $ is the mean relative velocity between PAH$^-$ and ions, which is given by the expression 
\begin{equation}
 \left <v_{0}\right > = \sqrt{\frac{8{\rm k_B}T}{\pi \mu}}, 
 \end{equation}
 with $T$ the temperature and $\mu$ the reduced mass, which is mainly determined by the mass of 
   X$^+$.  In expression (\ref {k_grain_mol}),  we have assumed that the radius of the PAH is the critical radius for which reaction can occur.  The  radius at which the second term equals the first term on the right-hand-side is approximately 
 1~$\mu$m at 10~K. This radius is much larger than the PAH radii considered here, so that $\frac{{\rm e}^{2}}{a{k_B}T} >> 1$.  The rate coefficient at 10 K is approximately $2 \times 10^{-7}$ cm$^{3}$ s$^{-1}$ for an ion of 30 amu and a PAH of 4~$\AA$. 
 
 Relation~(\ref{k_grain_mol}) is based on the fact that, due to the delocalization of the negative charge on the PAH, the trajectory of the positive ion as it approaches the PAH will always be affected by this charge.  Taking into account the complex shape and large number of possible PAH clusters in dense clouds, several negative charges may stick, and the attraction between such a system and  a positive ion may depend on the localization of the electrons in the cluster. As a consequence,  expression~(\ref{k_grain_mol}) may be more likely to be correct for small PAH's of 4$\AA$ radius than for larger ones. 
For this reason, we  consider the alternative possibility that  rate coefficients for the reactions between $\rm PAH^-$ and X$^+$ are given by the simple ballistic  relation 
\begin{equation}\label{k_ball}
k  = \pi a^2 \left <v_{0}\right >. 
\end{equation}
For small PAH's, the rate coefficient varies as $a$ whereas for large PAH's, if the process is ballistic, it varies as $a^2$.  For a 100 $\AA$ PAH and an ion of 30 amu, the rate coefficient at 10 K is then approximately $2 \times 10^{-8}$ cm$^{3}$ s$^{-1}$, or 0.1 of the value for the smaller PAH's.

In addition to these reactions, PAH$^-$ is neutralized by photodetachment of electrons due to ultraviolet, visible, and even near infra-red photons:
\begin{equation}
\rm PAH^- + h\nu \longrightarrow  PAH + e^{-}. 
\end{equation}
Following \cite{1988ApJ...329..418L}, we assume a cross section of 10$^{-16}$ cm$^{2}$ significantly above threshold, which is the electron affinity \citep[see also][]{2007CP....332..353M}.  For a standard interstellar radiation field, this leads to a conservative estimate for the rate coefficient in the absence of shielding of approximately 10$^{-7}$ s$^{-1}$, neglecting differences in electron affinities among PAH's.   A crude estimate of the dependence on $A_{\rm v}$ is given by the simple exponent  $\exp(-A_{\rm v})$.  Even with an $A_{\rm v}$ of 10, the rate is not negligible  in dense clouds.  Indeed, under most conditions, photodetachment is more rapid than neutralization, but it must be remembered also that the electrons produced in photodetachment then attach to neutral PAH.
We do not consider photodetachment by cosmic ray-induced photons. If we assume a rate coefficient of $2\times 10^4 \times \zeta$ for this process \citep{2003MNRAS.343..390F}, the photodetachment by direct UV photons is then more efficient by one order of magnitude even at an $\rm A_v$ of 10.

 In our original calculations, we did not include PAH$^{+}$.  In dense clouds, positively charged PAH's can be produced by secondary UV photons induced by cosmic-rays and by charge transfer with small positive ions X$^{+}$. The latter process can neutralize the X$^{+}$ abundance and so affect the small-molecule chemistry.  Then  PAH$^+$ can be destroyed by recombination with free electrons.  As suggested by the referee, we tested the importance of positively-charged PAH species using rate coefficients from \citet{2003MNRAS.343..390F} for PAH$^+$ production by photoionization and for recombination with electrons. For the charge-transfer reactions, we used rate coefficients computed with the Langevin theory and the polarizability from \citet{1986A&A...164..159O}. As an example, the Couloub-based rate coefficient for the recombination reaction between PAH$^-$ and X$^+ $ is $7.5\times 10^{-8}$ cm$^{3}$ s$^{-1}$ for PAH's of 4 $\AA$ and an ion of 30 amu whereas the rate coefficient for the charge-transfer reaction between PAH and X$^+$ for the same ion is $3.2\times 10^{-9}$ cm$^{3}$ s$^{-1}$.  In the various models that we ran, we found that  the abundance of PAH$^+$ is at least two orders of magnitude smaller than the PAH and PAH$^-$ abundances.  Moreover, its inclusion does not affect any of the results discussed below. This conclusion holds even when the recombination rate coefficient is the ballistic one. 

\section{Model parameters}\label{model_other}

We utilize a pseudo-time-dependent gas-phase model, which computes the chemical evolution at a fixed temperature and density. A version of this model (NAHOON) is now available online at \\ \verb+http://www.obs.u-bordeaux1.fr/radio/VWakelam/Valentine\%20Wakelam/Downloads.html+. The chemical database is a recent version of the OSU network: \verb+osu_01_2007+ , which can be found online at \verb+http://www.physics.ohio-state.edu/~eric/research.html+. The reactions involving PAH's will also be available at the same address.

We use three different sets of elemental abundances (EA1, EA2 and EA3), which are listed in Table~1 with respect to the total nuclear hydrogen abundance along with the initial choice of ionization.  EA1 represents the low-metal abundances discussed earlier.  EA2 is based on the  high-metal abundances observed in the $\zeta$ Ophiuchi diffuse cloud, but is modified based on recent observations. The abundance of helium has not been observed in cold cores and we use a value of 0.09 for EA2 with respect to hydrogen, based on observations in the Orion Nebula \citep{1991ApJ...374..580B,1992ApJ...389..305O}. To account for the oxygen and carbon that are depleted on grain mantles in the form of H$_2$O and CO, we removed 2.4\% of C and 13\% of O  \citep{1995A&A...296..779S}.  It is possible that the metals are more depleted in dense clouds than in diffuse regions. Thus, for EA3, we assumed that the silicon and magnesium are totally depleted on grains  and we assumed an abundance of $1.5\times 10^{-8}$ for iron as suggested by \citet{2003MNRAS.343..390F}. The complete depletion of Si is one explanation for the non-detection of SiO in dense clouds \citep[see][]{1989ApJ...341..857Z}. We also considered an additional depletion of a factor of 10 for chlorine as suggested by the observation of HCl in cold material around OMC-1 \citep{1985ApJ...295..501B}. 
Note that in all the elemental abundances that we considered, the elemental C/O ratio is 0.4.

The species are assumed to be initially in atomic form except for hydrogen,  which is assumed to be entirely in its molecular form. We consider typical physical conditions for dense clouds: a temperature of 10~K, an H$_2$ density of $10^4$~cm$^{-3}$, a cosmic-ray ionization rate $\zeta$ of $1.3 \times 10^{-17}$~s$^{-1}$ and a visual extinction $A_{\rm v}$ of 10. 

\section{General effects of PAH's}\label{General_effect}

 In this section, we present results for the three sets of elemental abundances (EA1, EA2 and EA3) with and without PAH's. In the models with PAH's, referred to as EA/PAH, we assume their fractional abundance to be $3.07\times 10^{-7}$, our standard value for small PAH's.  Variations in this value will be considered in Sect. \ref{sensitivity} and \ref{comp_obs}.

\subsection{Ionization fraction}

\begin{figure}
\plotone{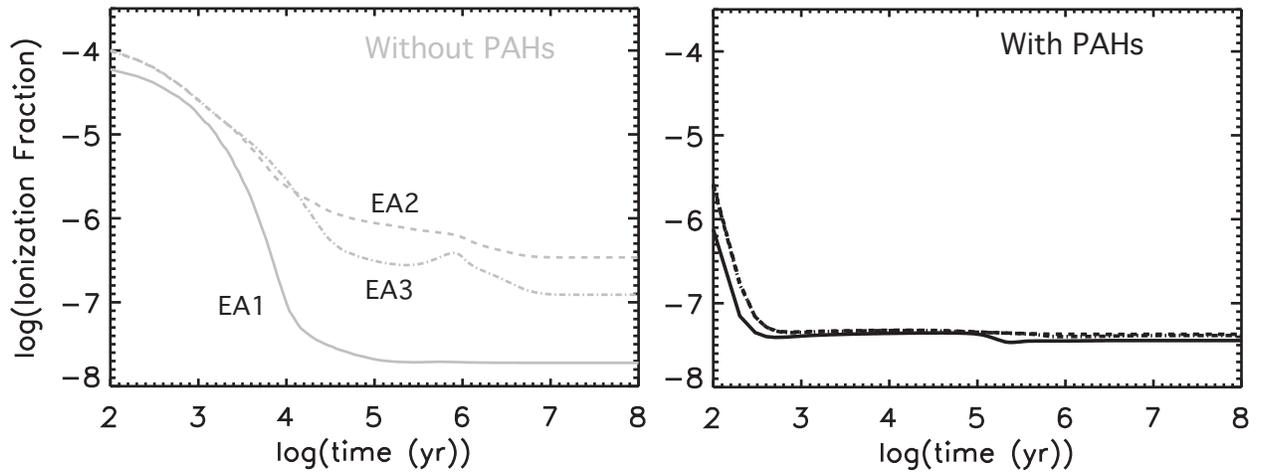}
\caption{Ionization fraction as a function of time for the three elemental abundances (solid line: EA1; dashed line: EA2; dashed-dotted line: EA3) without PAH's (grey lines; left panel) and with PAH's (black lines; right panel). } 
 \label{fract_ion}
\end{figure}

 Figure~\ref{fract_ion} shows the ionization fraction (sum of the negatively  or positively charged species) as a function of time for the three sets of elemental abundances, each with and without PAH's. The initial ionization is typical of a diffuse cloud since we start with ionized carbon in all models.  With low-metal elemental abundances (EA1) and no PAH's, the ionization fraction stays high for only $10^3$~yr and then decreases sharply, reaching a steady-state value of $3 \times 10^{-8}$ by 10$^{5}$ yr.  With both EA2 and EA3, and no PAH's, the ionization fraction decreases slowly, remaining larger than $10^{-6}$  for about $10^4$~yr and eventually reaching a steady-state value of a few $10^{-7}$.  When PAH's are added, the ionization fraction does not depend significantly on the elemental abundances. The main effect is to sharply decrease the earlier ionization fraction of the gas: within  the first 300~yr,  the ionization fraction goes from $\sim 10^{-4}$ to $\sim 5\times 10^{-8}$, its steady-state value, which is only slightly larger than that achieved with the EA1 model.  Thus, the inclusion of PAH's reduces the ionization fraction for EA2 and EA3 to a value close to that achieved with low-metal abundances and no PAH's. The steady-state ionization fraction for model EA1/PAH is larger by a factor of 2 compared with model EA1 without PAH's. The ionization fraction of model EA2/PAH at steady-state can be compared with the very similar value of $3.8 \times 10^{-8}$ reported by \cite{1988ApJ...324..553L} for a ratio of total PAH to hydrogen of $10^{-7}$ and high-metal abundances.  

\begin{figure}
\plotone{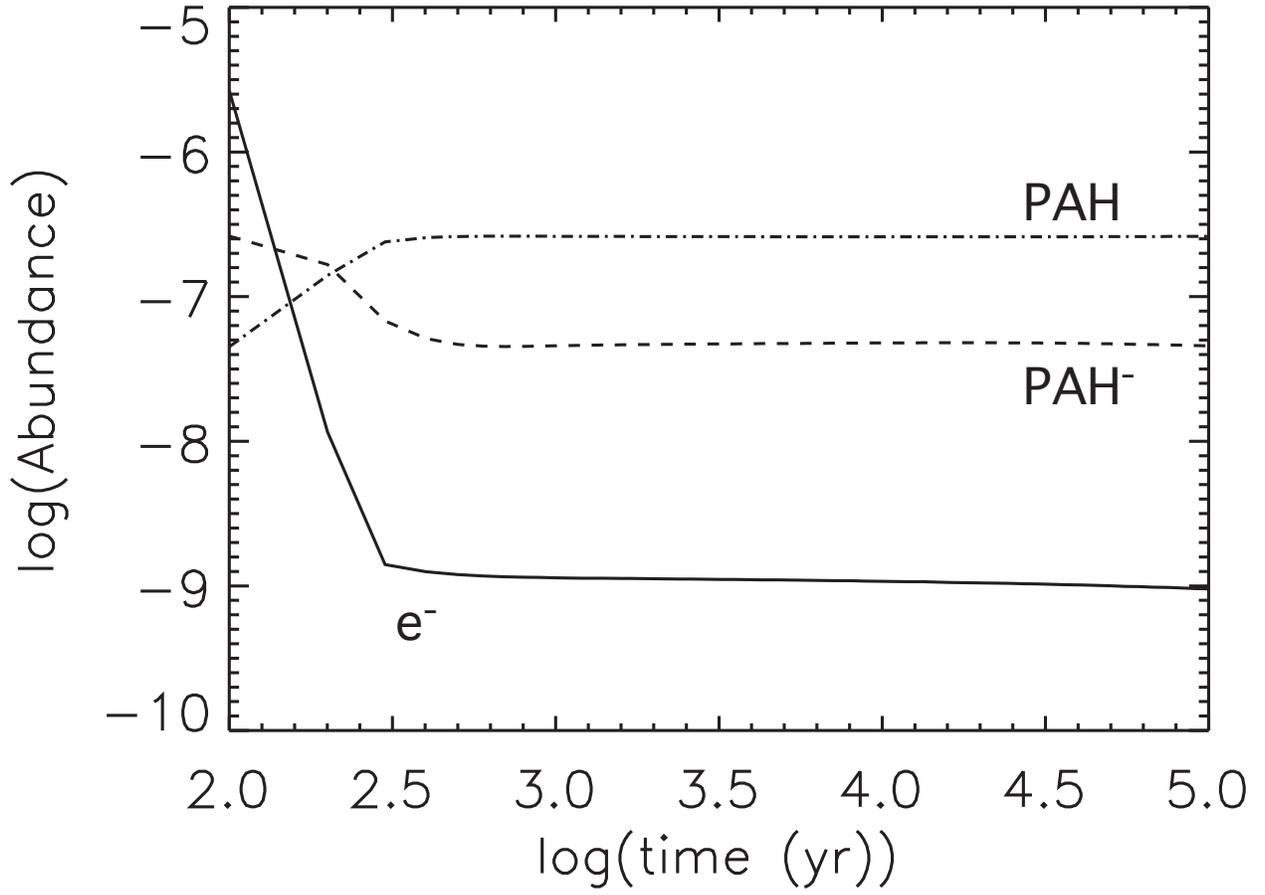}
\caption{Electron, PAH$^-$ and PAH abundances plotted vs.  time for the model EA2/PAH. } 
 \label{E_PAH}
\end{figure}

 Figure~\ref{E_PAH} depicts the fractional abundances of neutral and negatively-charged PAH's as well as the electron abundance as functions of time for model EA2/PAH. The other PAH models are not shown because these abundances do not depend significantly on the chosen elemental abundances.   By 300~yr, the fractional abundances of electrons and PAH$^-$ drop to $10^{-9}$ and $\sim 7\times 10^{-8}$ respectively, so that PAH$^-$ becomes the dominant 
  negatively-charged species. The conversion of neutral to negatively-charged PAH's occurs via electron attachment, which is very efficient at early times because of the high electron abundance.  But, once formed,  PAH$^-$ is efficiently destroyed by reactions with atomic ions, which are also very abundant  at the beginning of the chemical evolution. Note that the recombination of atomic ions is much more efficient with PAH's than with electrons since the latter must occur radiatively. For instance, the neutralization of C$^+$ by PAH$^-$ (with $a=4 \AA$) is four orders of magnitude larger than by electrons at 10~K. Because of  this efficient neutralization, the ionization fraction of the gas decreases much more quickly when PAH's are present (see Fig. ~\ref{fract_ion}).   We also find that the abundance of neutral PAH's exceeds that of the anionic PAH's by a factor of 5-10 after 300 yr, which is also in agreement with the steady-state results of \cite{1988ApJ...324..553L}.

\subsection{Results for selected species}

In the absence of PAH's, the higher-metal elemental abundances  EA2 and EA3 lead to a very different chemistry from EA1, and tend to worsen the agreement between the modeling results and the observations for selected classes of molecules  \citep{1982ApJS...48..321G}. With models EA2/PAH and EA3/PAH, however, the chemistry is much closer to that computed with the low metal elemental abundances in the absence of PAH's. The major change in the chemistry is caused by the high abundance of PAH$^{-}$, which neutralizes atomic positive ions efficiently and, upon reaction with positive molecular ions,  tends to neutralize them without significant dissociation, contrary to dissociative recombination reactions with electrons. In addition, the decrease of the ionization fraction at initial stages of the chemistry  increases the abundance of ionic species such as H$_3^+$ and H$_3$O$^+$ because they are depleted more slowly. At later times, some of the C-, S-, N-bearing species are enhanced in the models EA2/PAH and EA3/PAH because of i) larger elemental abundances and ii) a smaller electron abundance.  We now present the results for the most important species and for selected molecules that show significant differences among models.  

\subsubsection{C, C$^+$ and CO}

\begin{figure}
\plotone{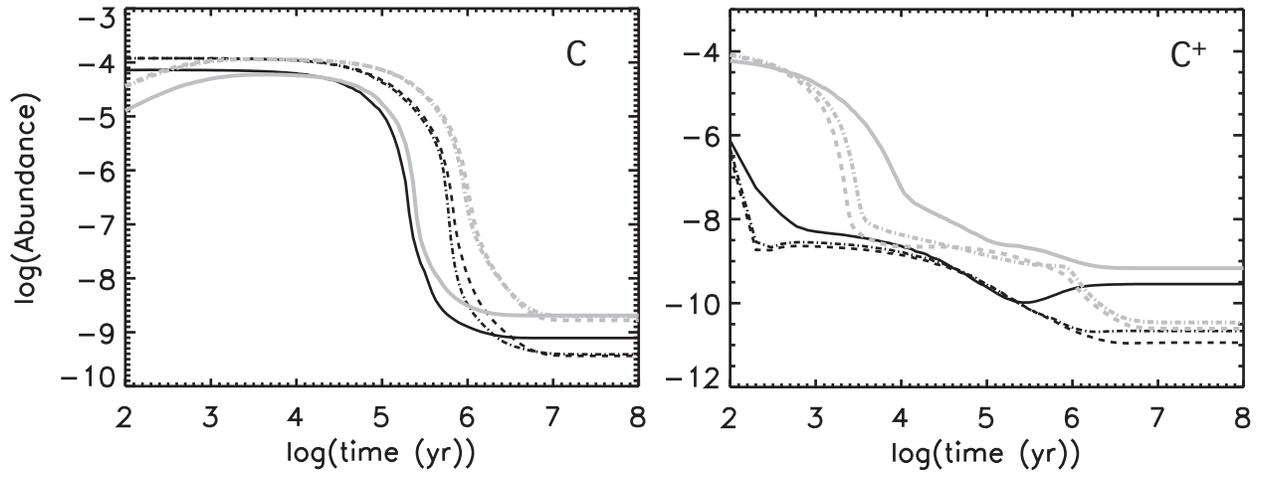}
\caption{C and C$^+$ abundances as a function of time for the three elemental abundances (solid line: EA1; dashed line: EA2; dashed-dotted line: EA3) without PAH's (grey lines) and with PAH's (black lines).  } 
\label{C_fam}
\end{figure}

In Figure \ref{C_fam}, we show our results for C and C$^+$  as functions of time for the three elemental abundances with and without PAH's. As expected, the fractional abundance of ionized carbon is decreased most rapidly when the PAH's are included.  The agreement at steady-state for the model EA2/PAH is in reasonable agreement with the result of \cite{1988ApJ...324..553L}.
A consequence of the C$^+$ neutralization by PAH$^-$ is that  atomic carbon becomes the main reservoir of carbon earlier than with the models without PAH's.  At the age of dense clouds, between $10^5$ and $10^6$~yr, atomic carbon is more abundant in EA2 and EA3 compared with EA1 because of a higher elemental abundance of carbon.   Our steady-state fractional abundance of neutral carbon ($\sim 3 \times 10^{-10}$) in model EA2/PAH is three orders of magnitude below that of the analogous model in \citet{1988ApJ...324..553L}.  After early time, CO becomes the main repository of carbon in all models; the CO abundance (not shown in the figure) is insensitive to the model used.

\subsubsection{H$_3^+$, HCO$^+$, H$_3$O$^+$, OH, H$_2$O and O$_2$}\label{O_spec}

\begin{figure}
\plotone{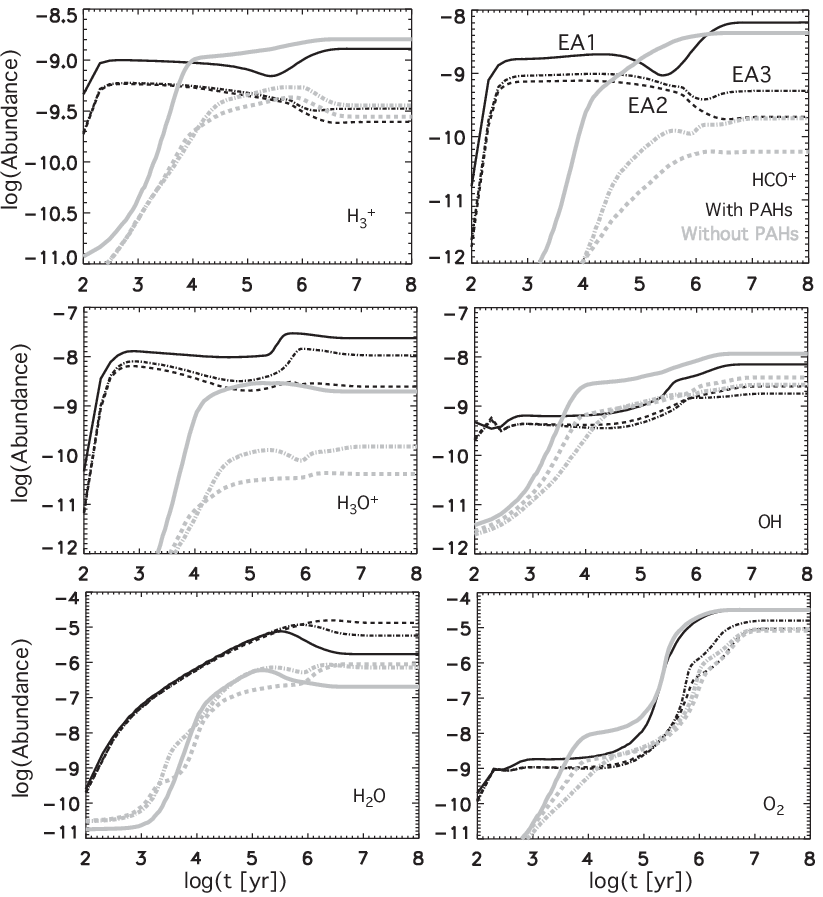}
\caption{H$_3^+$, HCO$^+$, H$_3$O$^+$, OH, H$_2$O and O$_2$ abundances as a function of time for the three elemental abundances (solid line: EA1; dashed line: EA2; dashed-dotted line: EA3) without PAH's (grey lines) and with PAH's (black lines).  } 
\label{OH_fam}
\end{figure}

The rapid decrease of the electron abundance in the models with PAH's leads to strong increases in the abundances of  molecular ions such as H$_3^+$, H$_3$O$^+$ and HCO$^+$ and neutral species formed from them at the initial stage of the chemistry,  as can be seen in Fig.~\ref{OH_fam}.  At longer times,  the ionic abundances show much less time dependence, with two  general but not universal patterns: (i)  they are more abundant in the models with PAH's than in the analogous models without them, and (ii) they are most abundant in models with EA1 abundances and least abundant in models with EA2 abundances, with and without PAH's.  General reasons for the trends include the higher electron abundances in models EA2 and EA3, the slower depletion rates by negatively charged species (mainly PAH$^{-}$) in the PAH-containing models, and differences in abundances of important neutral species involved in depletion processes among the PAH-containing models.

The ion H$_{3}^{+}$ is a partial exception to the trends in that  its abundance is not enhanced by PAH's at later times, mainly because it is destroyed principally by heavy neutral species such as CO and H$_{2}$O rather than by charged species.   The abundance in EA2 and EA3 is smaller than in EA1 because of dissociative recombination by more abundant electrons whereas it is smaller in EA2/PAH and EA3/PAH than in EA1/PAH because of reaction with H$_2$O, which is more abundant in the high-metal models. The  enhancement in the ion HCO$^{+}$ for models with PAH's in the initial stages of the chemistry is even more dramatic than that of H$_{3}^{+}$ because of its formation from abundant carbon atoms, by reactions leading first to CH$_{3}^+$.
At later times, HCO$^+$ is formed by the reaction between H$_3^+$ and CO in all models. Without PAH's, HCO$^+$ is destroyed by electrons whereas it is destroyed by H$_2$O in models with PAH's. 

The abundance of  H$_{3}$O$^{+}$ is also dramatically enhanced in the initial stages of the chemistry in the presence of PAH's.  Without PAH's, H$_3$O$^+$ is destroyed by electrons whereas it is destroyed by PAH$^-$ in EA1/PAH and EA3/PAH. In EA2/PAH, due to a large abundance of elemental silicon, H$_3$O$^+$ reacts with gas phase SiO, which is as abundant as $1.5\times 10^{-6}$ at $10^6$~yr.  Such a large abundance of SiO is  in disagreement with observations in dense clouds \citep{1989ApJ...341..857Z}. 
  
One consequence of the large  H$_3$O$^+$ abundance, at initial stages of the chemistry,  is that the abundances of OH, H$_2$O and O$_2$, which stem from the recombination and neutralization of H$_{3}$O$^{+}$,  are also enhanced in PAH-containing models .  The situation is somewhat more complex at later times with only water showing a much higher abundance in PAH-containing models. 
The predicted enhancement of H$_2$O in PAH-containing models is a problem because the calculated abundances far exceed the  observational upper limits of $\approx 3\times 10^{-7}$ in L134N and $7\times 10^{-8}$ in TMC-1 \citep{2000ApJ...539L.101S}.  The result for molecular oxygen is, on the contrary, not affected by PAH's at times after 10$^{5}$ yr.  Our theoretical O$_2$ abundance is about $10^{-9}$ at $10^5$~yr using EA2/PAH or EA3/PAH, which compares favorably with the observational upper limits of $1.7\times 10^{-7}$  in L134N and $7.7\times 10^{-8}$ in TMC-1 \citep{2003A&A...402L..77P}. 
 
\subsubsection{Assorted Hydrocarbons and Oxygen-containing Organic Molecules}

\begin{figure}
\plotone{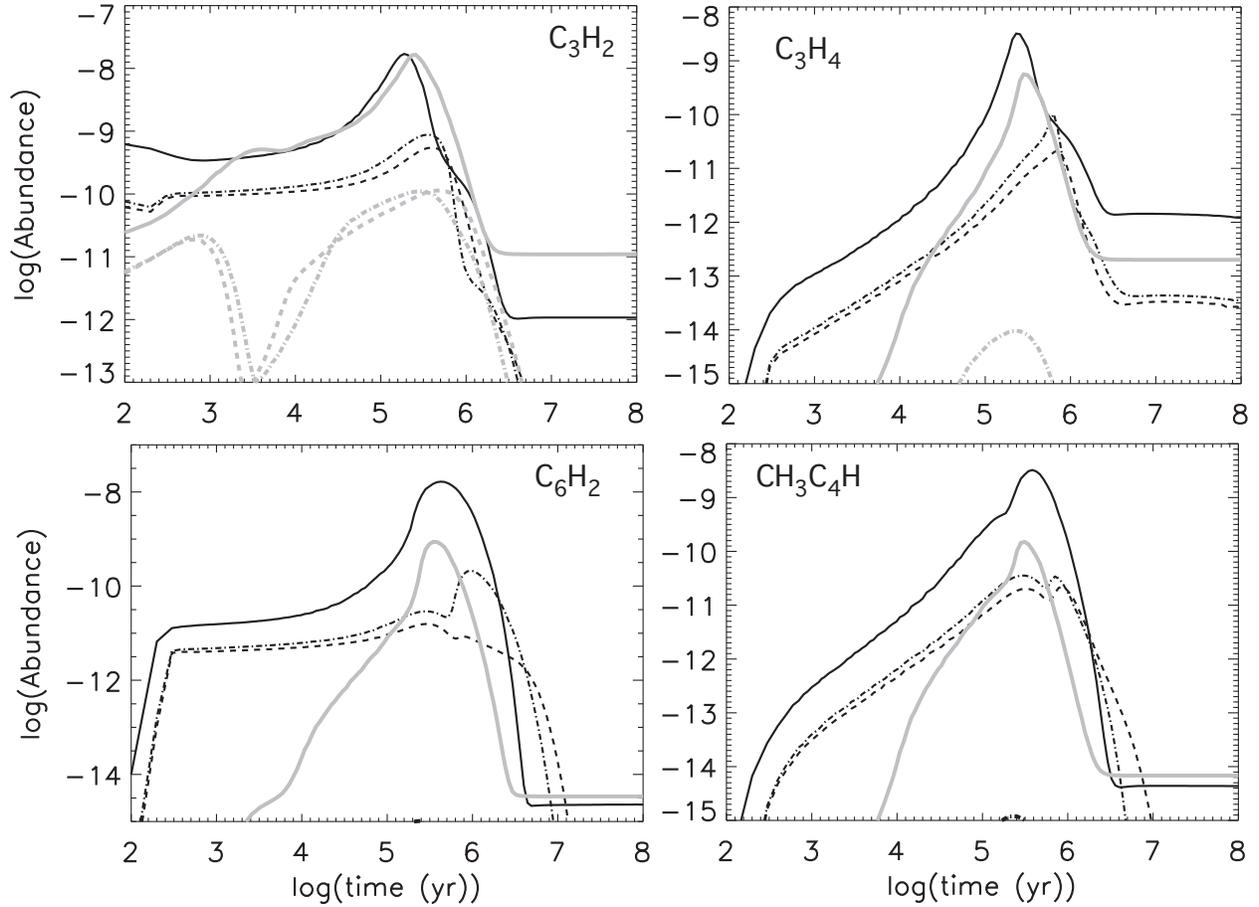}
\caption{c-C$_3$H$_2$ (cyclopropenylidene), C$_3$H$_4$ (methyl acetytene), C$_6$H$_2$ (triacetylene) and CH$_3$C$_4$H (methyl diacetylene) abundances as a function of time for EA1 (solid lines), EA2 (dashed lines), and EA3 (dashed-dotted lines) without PAH's (grey lines) and with PAH's (black lines).  For the three last molecules, the abundance produced by model EA2 and EA3 without PAH's is smaller than $10^{-15}$.  } \label{hydro_fam}
\end{figure}

The computed abundances of the hydrocarbons c-C$_3$H$_2$ (cyclopropenylidene), C$_3$H$_4$ (methyl acetylene), C$_6$H$_2$ (triacetylene) and CH$_3$C$_4$H (methyl diacetylene) are shown in Fig.~\ref{hydro_fam} as functions of time.  Without PAH's, the abundances peak at so-called early-times (10$^{5}$ yr) for low-metal abundances (EA1), a well-known effect.  For EA2 and EA3, on the other hand, the early-time peaks are weak to non-existent and the peak abundances very low, an effect caused by the high electron abundances, which disrupt the growth of molecular ions.  When PAH's are added, the abundances of the four hydrocarbons depicted are enhanced for all three models.  Although the enhancement is significantly greater for EA2 and EA3, the peak EA1 values are still the largest.  The enhancements are caused mainly by the lesser destructive effect of PAH$^{-}$ compared with electrons when it reacts with intermediate ions in the chain of synthetic reactions, although in individual cases secondary effects can be isolated.   The major effect, which comprises both lower rates and larger product neutral channels,  is stronger than the enhancement in destruction rates of hydrocarbons due to often larger abundances of molecular ions.  That differences remain among the PAH models despite the fact that PAH$^{-}$ has the same abundance shows, once again, that other processes are also important.
The computed abundances of hydrocarbons  at steady-state can be very different from  the results of 
\cite{1988ApJ...324..553L}. For example, the C$_3$H$_2$ abundance is $\sim 10^{-15}$ in model EA2/PAH  whereas it is $\sim 10^{-9}$ in \cite{1988ApJ...324..553L}.  This difference may be due to discrepancies in the network for carbon chemistry.

\begin{figure}
\plotone{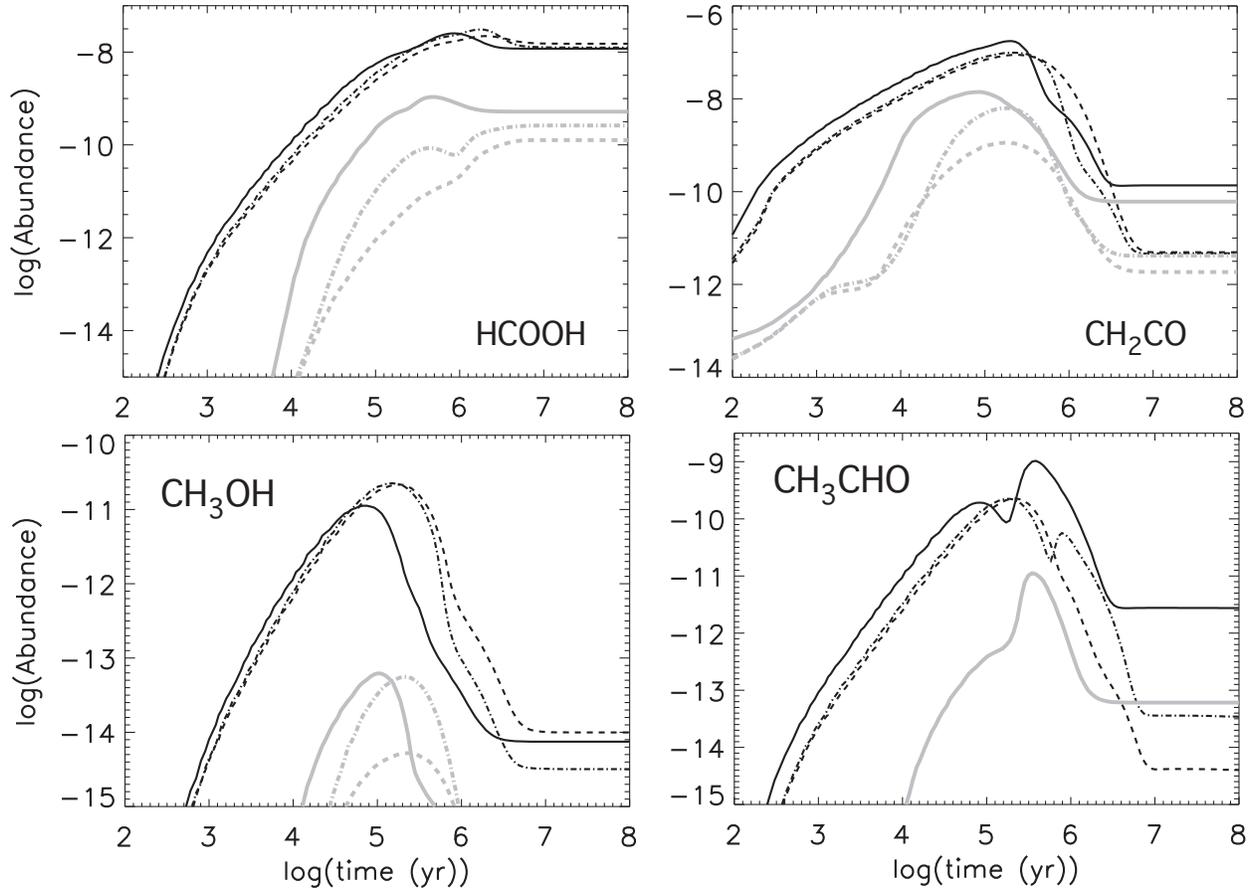}
\caption{The abundances of formic acid (HCOOH), ketene (CH$_2$CO),  methanol (CH$_3$OH), and acetaldehyde (CH$_3$CHO) are plotted as functions of time for three elemental abundances (solid line: EA1; dashed line: EA2; dashed-dotted line: EA3) without PAH's (grey lines) and with PAH's (black lines). The acetaldehyde abundance in model EA2 and EA3 is smaller than $10^{-15}$. } 
\label{CH_fam} 
\end{figure}

 As can be seen in Fig.~\ref{CH_fam}, the abundances of the oxygen-containing organic neutral molecules HCOOH, CH$_2$CO, CH$_3$OH and CH$_3$CHO in models containing PAH's are enhanced at almost all times by at least one order of magnitude over the abundances in models without PAH's.  Once again, the reason is generally the same as for the hydrocarbons: neutralization reactions between precursor ions with PAH$^{-}$ are not as destructive as dissociative recombination reactions with electrons.   Even if a species, such as ketene, is formed by a neutral-neutral reaction, its precursor is likely formed at least partially  through ion-molecule channels.  An as example, consider the case of methanol.  Methanol (CH$_{3}$OH) is produced through the ion CH$_{3}$OH$_{2}^{+}$, which itself is formed via the radiative association between CH$_{3}^{+}$ and water.  In models without PAH's, the protonated ion can recombine dissociatively with electrons to form CH$_{3}$OH + H, although this channel is now known to be a minor one \citep{2006GeppertFarad}.  With PAH$^{-}$, on the other hand, we assume that this channel is the only one, since the neutral form CH$_{3}$OH$_{2}$ does not exist and we only allow one hydrogen to be ejected from the species (see section~\ref{PAH_chem}).   Another reason for the enhanced abundance of methanol lies in the enhanced abundance of water.   We note that with PAH's, the peak predicted fractional abundance of methanol is still 1-2 orders of magnitude below the typical value in cold dense clouds of 10$^{-9}$ \citep{2006Garrodc}.
 
Unlike the situation with the hydrocarbons,  the abundances in models EA2/PAH and EA3/PAH are not significantly less and sometimes even greater than those in model EA1/PAH.  The small differences suggest that PAH$^{-}$ plays an even more dominant role than in the case of the hydrocarbons. In the non-PAH models, although the EA1 abundances tend to be greater than those in EA2 and EA3, the effect is smaller than for hydrocarbons except for acetaldehyde.  

\subsubsection{Cyanopolyynes}\label{HCmN}

\begin{figure}
\plotone{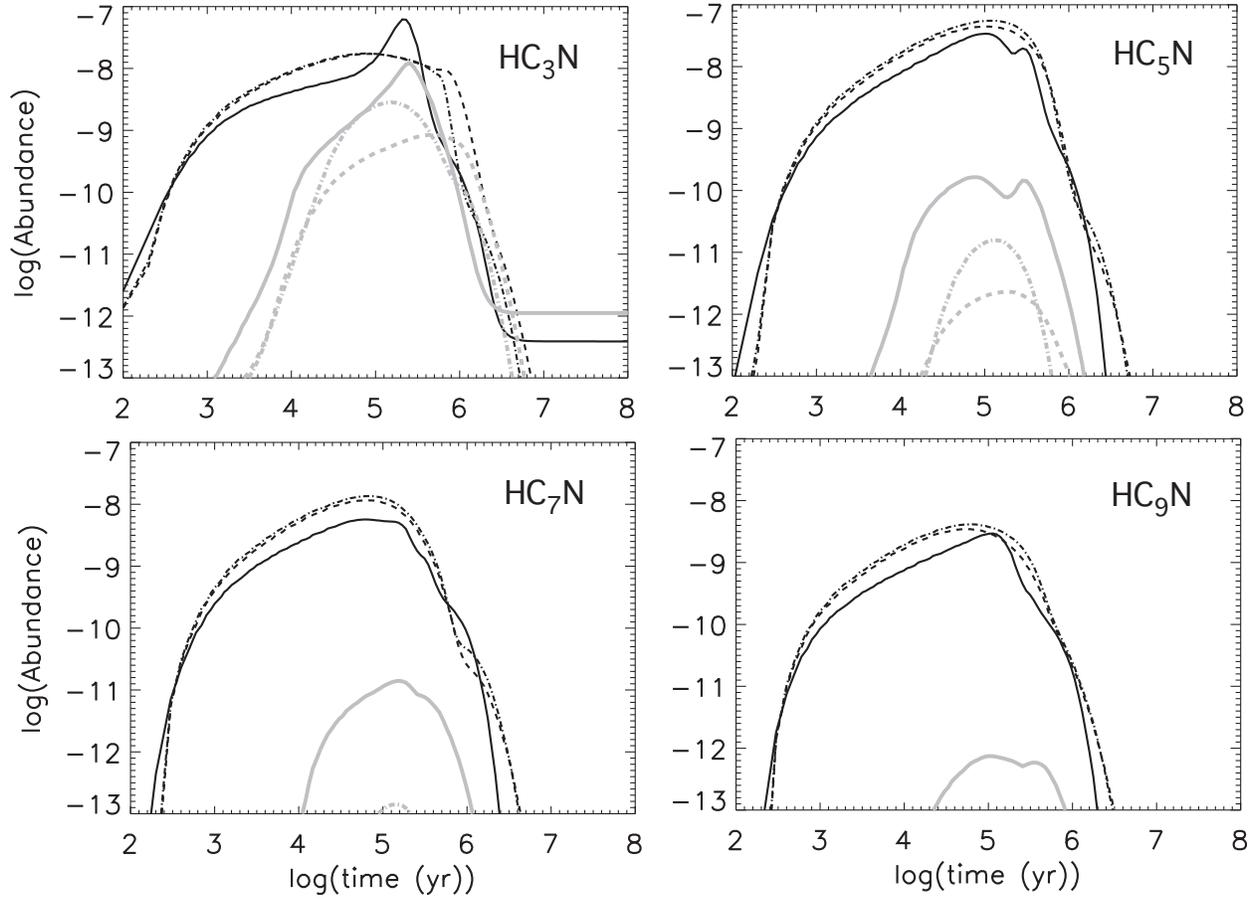}
\caption{Cyanopolyyne abundances as a function of time for the three elemental abundances (solid line: EA1; dashed line: EA2; dashed-dotted line: EA3) without PAH's (grey lines) and with PAH's (black lines).  The HC$_7$N and HC$_9$N abundances with models EA2 and EA3 are too small to be seen.} 
 \label{HCN_fam}
\end{figure}

Cyanopolyynes are very abundant towards the famous cyanopolyyne peak (CP) of the cold core TMC1, yet their abundances are under produced  by gas-phase models unless a high C/O elemental abundance ratio is assumed \citep{1998ApJ...501..207T}.  As shown in Fig.~\ref{HCN_fam}, the PAH's have a strong enhancing effect on the formation of these molecules, especially on the more complex cyanopolyynes.  Once again, there is little difference among the results for the three PAH models.  In general, the dominant formation mechanism for each cyanopolyyne more complex than HC$_{3}$N involves the next smaller member of the series.  At $10^5$~yr, the HC$_5$N, HC$_7$N and HC$_9$N abundances are enhanced by more than three orders of magnitude over their values without PAH's.  The peak fractional abundances are indeed larger than the values observed in TMC1 (CP) of  $2\times 10^{-9}$, $5\times 10^{-10}$ and $2.5\times 10^{-10}$  for HC$_5$N, HC$_7$N and HC$_9$N respectively \citep[see][and references therein]{2000ApJ...535..256M}.  Similar enhancements are seen for other nitrogen-containing molecules such as CH$_{2}$CHCN, CH$_3$CN, CH$_3$C$_3$N and C$_3$H$_2$N$^+$. 

\subsubsection{sulfur-bearing species}

\begin{figure}
\plotone{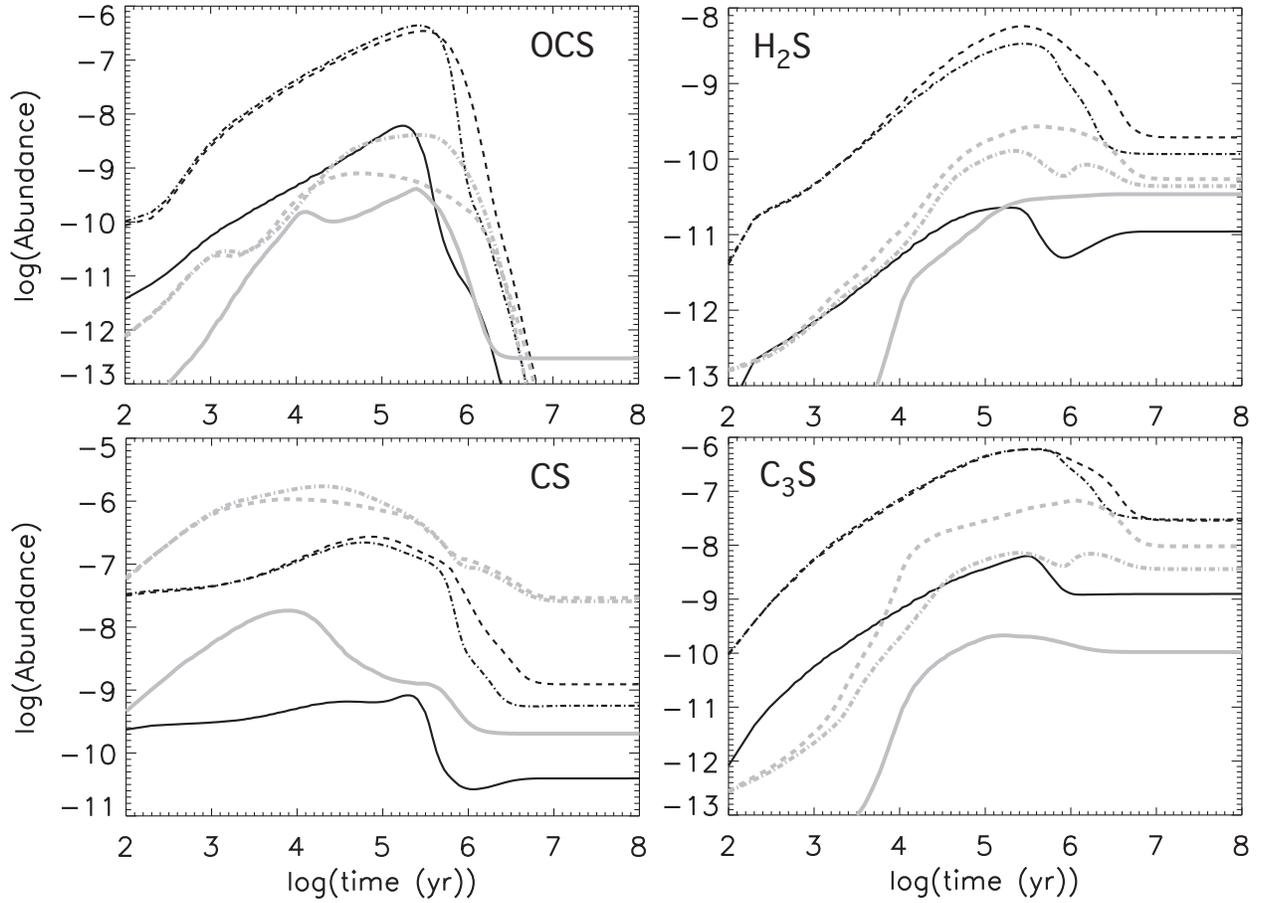}
\caption{The abundances of OCS, H$_2$S, CS and C$_3$S as  functions  of time for the three elemental abundances (solid line: EA1; dashed line: EA2; dashed-dotted line: EA3) without PAH's (grey lines) and with PAH's (black lines). } 
 \label{S_fam}
\end{figure}

The elemental abundance of sulfur is smaller by more than 2 orders of magnitude in EA1 compared with the two other sets of elemental abundances. 
As a consequence, most of the sulfur-bearing species have larger abundances with high-metal abundances at all times, with and without PAH's.  In Fig.~\ref{S_fam}, we follow the temporal evolution of four species: OCS, H$_2$S, CS and C$_3$S.  
Three of these molecules tend to have enhanced peak abundances in the presence of PAH's, although the situation is more complex towards steady-state.  The exceptional cases are CS for all elemental abundances and H$_{2}$S for the low-metal case.  The CS molecule  is produced by the dissociative recombination of HCS$^+$ and HC$_2$H$^+$ by electrons  or dissociative neutralization of HCS$^+$ by PAH$^-$. The larger ion HC$_2$H$^+$ does not react with PAH$^{-}$ to produce CS with our assumptions. As a consequence, the abundance of CS is smaller in the models including the PAH's.  The late-time situation regarding H$_{2}$S is more complex. The primary formation mechanism for H$_{2}$S at most times is through the protonated precursor H$_{3}$S$^{+}$, which is formed by a chain of reactions starting with S + H$_{3}^{+}$ and proceeding through HS$^{+}$ and H$_{2}$S$^{+}$. Without PAH's, H$_2$S is  also produced  by the charge exchange between H$_2$S$^+$ and S. The ion H$_2$S$^+$ is more abundant in the low-metal case without PAH's because it is produced by the reaction $\rm S^+ + H_2CO \longrightarrow H_2S^+ + CO$ where S$^+$ is more abundant than the neutral form in the absence of PAH's. 

\section{Sensitivity to  several PAH parameters}\label{sensitivity}

\begin{figure}
\plotone{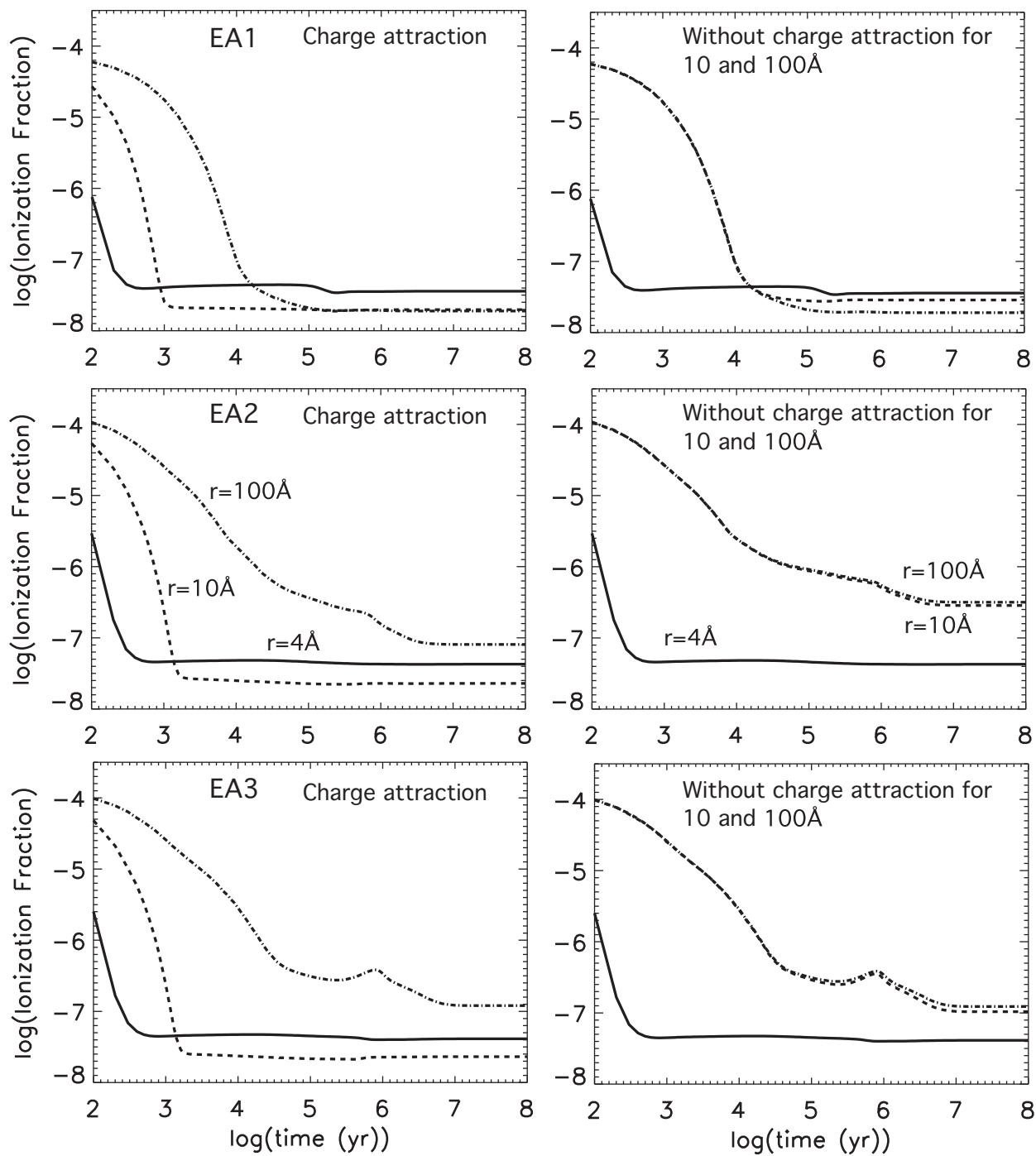}
\caption{Ionization fraction as a function of time for small  and large PAH's using  the three elemental abundances. For the two larger PAH's, we use expressions for the rate coefficients of PAH$^{-}$-X$^{+}$ neutralization reactions that include (left column) and exclude charge attraction (right column). } 
 \label{fract_othermod}
\end{figure}

There are some observational and theoretical indications that PAH's  in dense clouds may be aggregated into clusters with radius up to 100~$\AA$.   We have performed additional calculations using  the three elemental abundances and PAH-clusters of radii 10 $\AA$ and 100 $\AA$  with the reduced abundances discussed in Sect.~\ref{PAH_model}.  In our discussion of the results, we use the word PAH whatever the size.   For the larger PAH's (10 and 100 $\AA$), we consider two expressions for the rate coefficients of the neutralization reactions between PAH$^{-}$ and  positive ions: one 
involving charge attraction, as expressed  in equation (\ref{k_grain_mol}), and the other a simple ballistic one, as expressed in equation (\ref{k_ball}).  The ballistic rate is considerably lower. For radii of 10 and 100 $\AA$ and a temperature of 10~K, the ballistic rate is, respectively,  about 1000 $\times$ and 10 $\times$ smaller than the rate involving charge attraction.  Another process with a rate coefficient that depends on the PAH size is the radiative attachment of electrons to neutral PAH's, which increases with increasing radius.  In Fig.~\ref{fract_othermod}, we show the ionization fraction as a function of time with each set of elemental abundances and PAH's of  4, 10, and 100 $\AA$.  As can be seen, the ionization fraction depends highly on the size of the PAH, and on whether or not charge attraction is considered.   We do not show the dependence on the total abundance of C atoms incorporated into PAH's, but a larger abundance implies a more rapid decrease of the ionization fraction to a smaller value at late times. 

In  the left panel of Fig.~\ref{fract_othermod},  neutralization proceeds with charge attraction for all three sizes of PAH's.  For the two smaller sizes of PAH's, the main negatively-charged species in the model remains PAH$^-$.  For the PAH's of 100 $\AA$ radius, the total abundance of PAH's is $2\times 10^{-11}$, much smaller than the abundance of electrons. The evolution of the ionization fraction is then closer to the case without PAH's, but the steady-state ionization fraction can still be somewhat lower. For the case with model EA1 and the largest PAH's, the ionization fraction is the same  with or without charge attraction and is not different from the model in which there are no PAH's at all.    As a consequence,  the 10 $\AA$ and (for EA2 and EA3) the 100 $\AA$ cases represent intermediate stages between the models with 4 $\AA$ PAH's and  those without PAH's.
The right-hand panel of Fig.~\ref{fract_othermod} shows the results when positive ion-PAH$^-$ neutralization occurs ballistically  for the larger PAH's.  In this case, the 10 $\AA$ and 100 $\AA$ cases give very similar results to the model without PAH's so that the effect of large PAH's at the chosen abundances  is negligible in dense cloud chemistry.  This effect occurs because the major depletion mechanism for PAH$^{-}$ is photodetachment rather than recombination with positive ions, so that lowered neutralization rates have the sole effect of decoupling PAH$^{-}$ from the ion-molecule chemistry. 

 The modification of the ionization fraction has consequences for the abundances of  species.  In general, PAH's of 100 $\AA$ (considering charge attraction or not) and of 10 $\AA$ without charge attraction lead to abundances similar to the models without PAH's.  For most of the species, PAH's of 10 $\AA$ with charge attraction give intermediate abundances between the case of very small 4 $\AA$ PAH's and the case without PAH's. Many species such as C, C$^+$ H$_3^+$, HCO$^+$ and OH show a difference smaller than a factor of two while using larger PAH's of 10 $\AA$ instead of smaller PAH's of 4~$\AA$. Some species are largely underproduced using the elemental abundances EA2 and EA3 without PAHs. This is the case for C$_3$H$_2$, C$_3$H$_4$, C$_6$H$_2$, CH$_3$C$_4$H, CH$_3$CHO and cyanoppolyynes. In the presence of 10 $\AA$ PAH's, these species are almost as abundant as with smaller PAH's. 
 
\section{Comparison with observations}\label{comp_obs}

 The objective of this section is to study the impact of PAH's on the comparison with observations in the two cold cores L134N (north peak) and TMC-1 (cyanopolyyne peak). These two clouds are known to show differences in the abundances of C-rich molecules that are sometimes explained by  differences in age or C/O ratio \citep{2006A&A...451..551W}. Here we explore the possibility that they are also explained by the amount and size of PAH's present.

\subsection{Method}

The observed abundances in L134N (north peak) and TMC-1 (cyanopolyyne peak) are listed in Tables 3 and 4 of \citet{2006Garrodc}. In addition to these species, comprising 42 molecules for L134N and 53 for TMC-1, we considered SiO,  for which only upper limits of $3.6\times 10^{-12}$ for L134N and $2.4\times 10^{-12}$ for TMC1 have been determined \citep{1989ApJ...343..201Z}.

In this comparison, we utilize the three different sets of elemental abundances listed in Table~1. Of all these elements, sulfur is the least constrained by
 abundances of gas-phase molecules gas-phase abundances so that the depleted fraction is uncertain \citep{1999MNRAS.306..691R}. PAH size and abundance are also poorly determined      parameters. Instead of fixing these parameters to previously used values, we try to give some trends by comparing observations with grids of models in which these three parameters are varied in addition to time. 

We computed 441 models by varying the abundance of elemental sulfur between $10^{-8}$ and $10^{-5}$, the abundance of PAH's between $10^{-10}$ and $10^{-7}$ and the size of PAH's between 4 and 400~$\AA$. Note that the largest values of PAH sizes and abundances used here are not in agreement with the relation $N_C = 460 (a/10 \AA)^3$ used in the first part of the article. Considering the little information available on the relation of size to number of carbons for PAHs, in particular for PAH clusters, we consider the PAH size and abundance as independent.  The agreement between present models and observations quantified using statistical comparisons, $\chi^2$ minimization for instance, is not good enough to give reasonable constraints on model parameters. For this reason, we did not use such methods but instead for each model and each time, we count the number of species with fractional abundance X reproduced by the model within one order of magnitude ($\rm X_{obs}/10 < X_{model} < X_{obs}\times 10$) where $\rm X_{obs}$ is the observed abundance and $\rm X_{model}$ the computed abundance. For those species for which only upper limits to their abundances are known and for the case of atomic carbon, for which only a lower limit is known \citep{1996A&A...311..282S}, we check that the predicted abundance is in agreement with the observed limit when this is the case, we consider that the species is reproduced. 

\subsection{Results}

\begin{figure}
\plotone{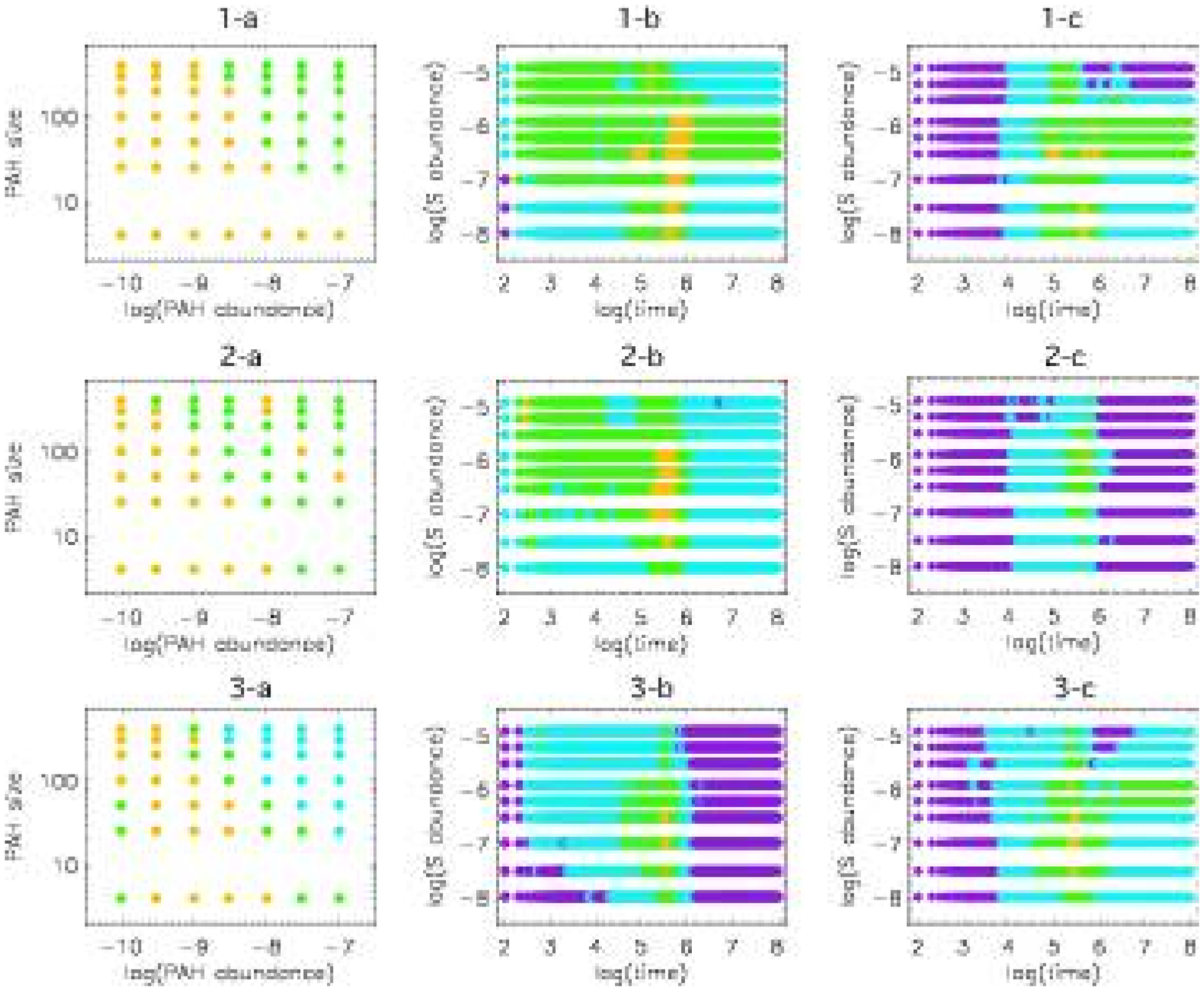}
\caption{Level of agreement (violet:  less than 50\%; blue, between 50 and 60\%; green, between 60 and 70\%; and yellow, between 70 and 80\%) between models and observations in L134N. Each row represents the results for a set of elemental abundances (1: EA1, 2: EA2 and 3: EA3). Column a represents the level of agreement as a function of PAH parameters with optimized sulfur and time at each point.  best agreements for sulfur elemental abundance and time as a function of PAH parameters. Column b represents the level of agreement as a function of the sulfur elemental abundance and time with optimized PAH parameters at each point. Column c represents the agreement without PAH's.   } 
 \label{L134N}
\end{figure}

\begin{figure}
\plotone{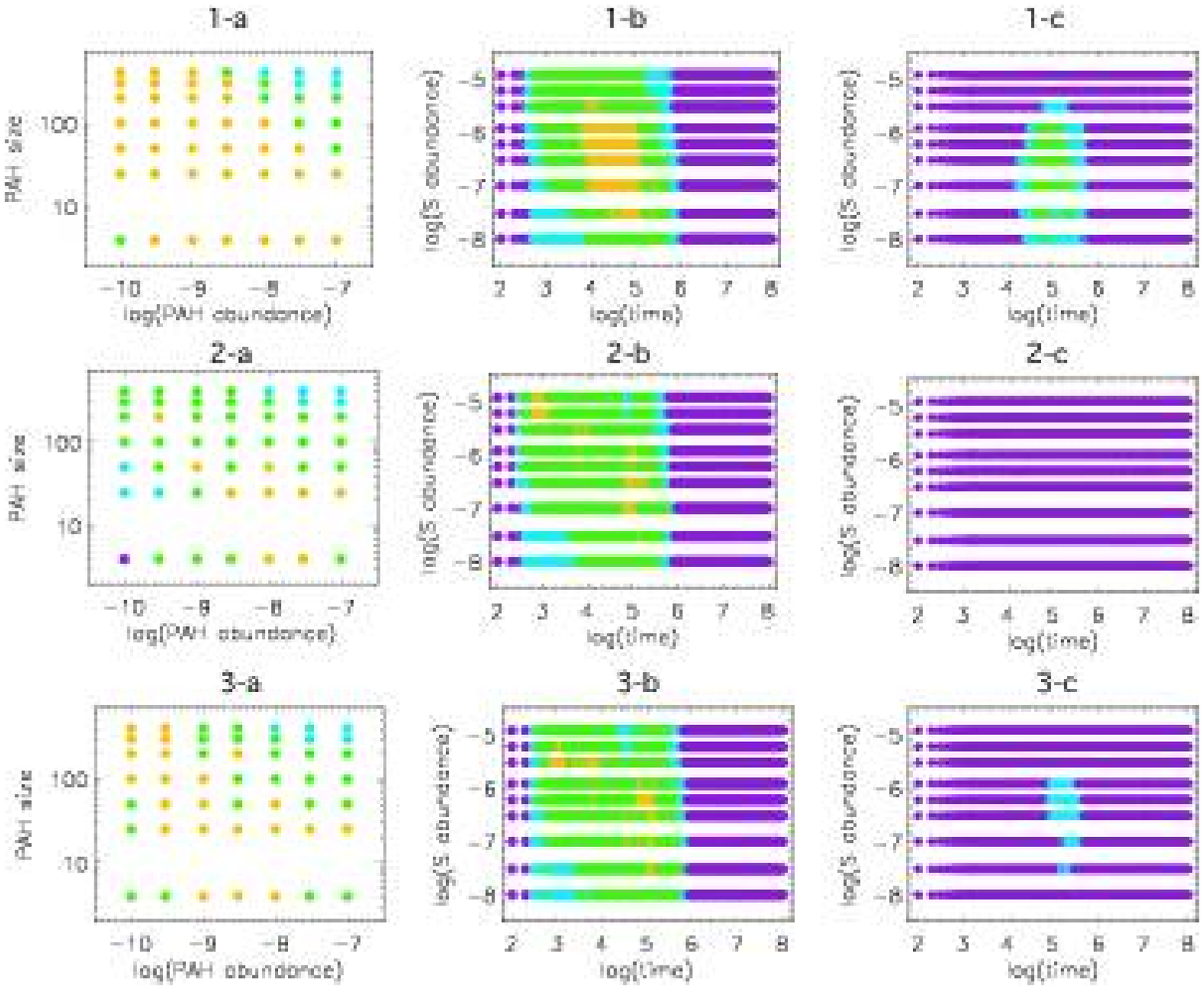}
\caption{Level of agreement (violet:  less than 50\%; blue, between 50 and 60\%; green, between 60 and 70\%; and yellow, between 70 and 80\%) between models and observations in TMC-1. Each row represents the results for a set of elemental abundances (1: EA1, 2: EA2 and 3: EA3). Column a represents the level of agreement as a function of PAH parameters with optimized sulfur and time at each point.  best agreements for sulfur elemental abundance and time as a function of PAH parameters. Column b represents the level of agreement as a function of the sulfur elemental abundance and time with optimized PAH parameters at each point. Column c represents the agreement without PAH's. } 
\label{TMC1} 
\end{figure}

Figures~\ref{L134N} and \ref{TMC1} show the maximum percentage of species reproduced in each cloud as a function of the parameters chosen. Column (a) shows the level of agreement as a function of PAH size and abundance, with each point optimized for the best elemental abundance of sulfur and the best time. Column (b) shows the level of agreement as a function of the elemental abundance of sulfur and the time optimized at each point for the best PAH parameters. Finally, column (c) shows the agreement as a function of the elemental abundance of sulfur and the time for models without PAH's. These results are shown for each set of elemental abundances (1: EA1, 2: EA2 and 3: EA3). Best agreement, with or without PAH's, is most often achieved at so-called early times of $10^{4-6}$ yr.  Moreover, high levels of agreement cannot be achieved with high abundances of PAH clusters, as shown in the upper-right corners in Column 1 for both figures.

The maximum percentage of molecules reproduced depends on all the parameters, and several solutions exist. PAH's do not improve significantly the best agreement for L134N (already above 70\% without PAH's) but they do expand the range of parameters (time and elemental abundance of sulfur) for which we have an agreement larger than 70\%, especially if elemental abundances 1 or 2 are utilized. Note particularly the expanded range of agreement with the use of elemental abundances 2 (the high-metal ones). The optimum PAH parameters corresponding to the yellow zones in Column b do not depend  on the optimum sulfur abundance.  As an example, high agreement is obtained with $3\times 10^{-10}$ abundance of 4$\AA$ PAH's both for low ($3\times 10^{-8}$) and high ($1\times 10^{-6}$) elemental abundance of sulfur. The chemistry in TMC-1 on the contrary is better reproduced if PAH's are present, as can be seen immediately by comparison of columns b and c in Figure \ref{TMC1}. Without PAH's, agreement rarely extends to more than 60\%, while with PAH's, agreement in excess of 70\% exists for all three sets of elemental abundances. For elemental abundances 1, the PAH parameters are relatively unconstrained; while for the high-metal abundances (EA2), high abundances of small PAH's are needed.  The situation is less clear for EA3. For both clouds, we find isolated areas of agreement larger than 70\% at very early times, $<1000$~yr, for large abundances of sulfur ($>8\times 10^{-6}$) and big ($>25 \AA$) PAH's ($>10^{-8}$) if EA2 and/or EA3 is utilized.  Such agreement is of mathematical interest only.

\section{Conclusion}

Most chemical studies of the dense interstellar medium rely purely on gas-phase processes.  A few contain, in addition, a varying amount of surface chemistry on interstellar grains but the inclusion of PAH's, which occupy a middle ground in size between small gas-phase species and dust particles, is quite rare.  Part of the reason for this neglect is that to include PAH's properly, one should view them as individual species.  But the actual individual species that make up the PAH content of interstellar clouds are not well constrained from spectral observations, and even their chemistry has not been studied in great detail.  So, inclusion of PAH's means that one must treat them as small grains, and follow just their charge state.  In this study, as in previous ones, we lump all PAH's into just three groups, depending on whether they are neutral, positively-charged, or negatively-charged.  Positively-charged PAH's were  initially assumed to be unimportant in cold dense sources, an assumption that was confirmed upon the request of the referee.  They were not included in the chemical network used for the results presented here. By including an assortment of chemical processes involving negatively-charged and neutral PAH's into our standard gas-phase network of reactions, we are able to determine the effect of PAH's on the time-dependent gas-phase chemistry of cold and dense interstellar cloud cores.  The gas-phase model employed is known as \verb+osu_01_2007+, and can be downloaded from the URL http://www.physics.ohio-state.edu/$\sim$eric/research.html.  To this model, we add electron attachment to neutral PAH's, neutralization between negatively-charged PAH's and positively-charged smaller ions, and photodetachment of negatively-charged PAH's by external photons.  Our standard model contains a fractional abundance of $3.07 \times 10^{-7}$ PAH's of size 4 $\AA$, corresponding to 30 carbon atoms.   In addition to this model, we vary the size and abundance of the PAH's to study the impact of these parameters on the chemistry.  We also use three different sets of elemental abundances with different abundances of heavy atoms.

In general, the dominant effect in our calculations is that electrons are replaced by PAH$^{-}$ as the dominant negative charge carrier, an effect which weakens and eventually vanishes as the size of the PAH's, due probably to clustering,  and their fractional abundance decrease. When the abundance of PAH$^{-}$ exceeds that of electrons,  the overall ionization fraction is reduced because atomic positive ions are far more rapidly neutralized by reactions with PAH$^{-}$ than by radiative recombination with electrons.  These changes lead to many secondary effects.  Our calculations show that many of the molecular species have abundances that are strongly affected by the presence of PAH's. In particular, molecular ions tend to have increased abundances at most times and large unsaturated neutrals, such as the polyynes, as well as saturated species such as methanol, show very large increases in abundance. The sensitivity of our results to variations in the size of PAH's depends upon assumptions made in the rate of neutralization of PAH$^{-}$ and positive ions. The differences among results obtained with different elemental abundances are much smaller in models containing PAH's than in models without them.

 By comparing observations in the two dense cloud cores L134N and TMC-1 with grids of models, we showed only that the chemistry in such sources is not consistent with large amounts of large PAH clusters. We are not able to give more severe unambiguous constraints on the PAH parameters, since high levels of agreement can exist for large parameter variations.  The situation is similar for the elemental abundance of sulfur, where values over several orders of magnitude are consistent with maximum agreement with observations. As regards time, both the standard early-time and very early times can lead to optimum agreement, although the latter must be rejected on astronomical grounds. The situations for L134N and for TMC-1 show differences.  For L134N, the best agreement is not improved when PAH's are included, although the range of optimum sulfur elemental abundances and times are extended, most significantly so for the high-metal abundances (EA 2).  For TMC-1, on the other hand,  the best level of agreement is significantly better when PAH's are included, using any of the three elemental abundances. This difference may indicate a difference in PAH properties between the two regions. In particular, for EA 2, optimum agreement in TMC-1 is obtained with smaller and more numerous PAH's, while for L134N, such parameters make matters worse. Finally, unlike the case of gas-phase models without PAH's, models including PAH's are able to accommodate high-metal elemental abundances, similar to those observed in the diffuse medium, if the effects of PAH's are included and optimized.





\acknowledgments

V. W. acknowledges partial financial support from the CNRS/INSU program PCMI. E. H. thanks the National Science Foundation for its partial support of this work.

\clearpage

\begin{table}
\tablecaption{Elemental abundances with respect to $n_{\rm H}$. \label{elem_ab}}
\begin{tabular}{l|l|l|l|l}
\tableline\tableline
Species & EA1\tablenotemark{1} & EA2 & EA3 & 
Reference for EA2\\
\tableline
He & 1.40(-1) & 9.00(-2) & 9.00(-2) & see text \\
N & 2.14(-5) & 7.60(-5) & 7.60(-5) & \citet{1993ApJ...402L..17C}\\
O & 1.76(-4) & 2.56(-4)\tablenotemark{2} & 2.56(-4)\tablenotemark{2} & \citet{1998ApJ...493..222M} \\
C$^+$ & 7.30(-5) & 1.20(-4)\tablenotemark{2} & 1.20(-4)\tablenotemark{2} & \citet{1993ApJ...402L..17C}\\
S$^+$ & 8.00(-8) & 1.50(-5) & 1.50(-5) &  \citet{1993ApJ...413L..51F}\tablenotemark{3} \\
Si$^+$ & 8.00(-9) & 1.70(-6) & 0.00 & \citet{1994ApJ...420L..29C}\\
Fe$^+$ & 3.00(-9) & 2.00(-7)  & 1.50(-8) & \citet{1992ApJ...401..706S} \\
Na$^+$ & 2.00(-9) & 2.00(-7) & 2.00(-7) & \citet{Savage1996} \\
Mg$^+$ & 7.00(-9) & 2.40(-6) & 0.00 & \citet{1992ApJ...401..706S} \\ 
Cl$^+$ & 1.00(-9)  & 1.8(-7) & 1.80(-8) &  \citet{Savage1996}\tablenotemark{4}\\
P$^+$ & 2.00(-10) & 1.17(-7) & 1.17(-7) & \citet{Savage1996}\tablenotemark{4} \\
F$^+$ & 6.68(-9)\tablenotemark{5} & 1.8(-8) & 6.68(-9)\tablenotemark{5}  & \citet{2005ApJ...619..884F}\\
\tableline
\end{tabular}
\tablenotetext{ }{a(-b) refers to $\rm a\times 10^{-b}$}
\tablenotetext{1}{ Low-Metal abundances from \citet{1982ApJS...48..321G}; \citet{1974ApJ...193L..35M}}
\tablenotetext{2} {We considered an additional depletion compared to the observed value to take into account the oxygen and carbon depleted on grain mantles (see text).}
\tablenotetext{3} {The sulfur abundance found by the authors is larger than what is usually assumed as the cosmic abundance. We  use the cosmic abundance of $1.5\times 10^{-5}$.}
\tablenotetext{4}{ Abundances have been recalculated from the observations of  \citet{1974ApJ...193L..35M}. }
\tablenotetext{5} {Depleted value from \citet{2005ApJ...628..260N}.}

\end{table}%

 \clearpage

\end{document}